\documentclass{PoS}

\usepackage{amsmath}

\title{Lattice study on chiral dynamics of two-color six-flavors QCD}

\ShortTitle{Lattice study on chiral dynamics of two-color six-flavors QCD}

\author{\speaker{Masaaki Tomii}\\
        High Energy Accelerator Organization(KEK), Tsukuba 305-0801, Japan\\
        School of High Energy Accelerator Science, Graduate University for Advanced Studies (Sokendai),
        Tsukuba 305-0801, Japan}
\author{Masashi Hayakawa\\
        Department of Physics, Nagoya University, Nagoya 464-8602, Japan}
\author{Ken-Ichi Ishikawa\\
        Department of Physics, Hiroshima University, Higashi-Hiroshima 739-8526, Japan}
\author{Shinji Takeda\\
        School of Mathematics and Physics, College of Science and Engineering,
        Kanazawa University, Kanazawa 920-1192, Japan}
\author{Norikazu Yamada\\
        High Energy Accelerator Organization(KEK), Tsukuba 305-0801, Japan\\
        School of High Energy Accelerator Science, Graduate University for Advanced Studies (Sokendai),
        Tsukuba 305-0801, Japan
\\\\
        E-mail: \email{tomii@post.kek.jp}\\}

\abstract{
The electroweak symmetry breaking and origin of masses may be attributed to the breakdown of chiral symmetry due to a strong gauge dynamics. Among several candidates of such gauge systems, we focus on two-color QCD with $N_f=6$ massless Dirac fermions in the fundamental representation, and study on whether the dynamics of this gauge system trigger chiral symmetry breaking or not by simulating with Wilson fermions on lattices up to $L/a=32$. We show the result for the quark mass dependence and the volume dependence of some quantities such as the mass of the lightest pseudoscalar meson, decay constant, and give the three evidences supporting the absence of chiral symmetry breaking in the six-flavor theory.
}

\FullConference{31st International Symposium on Lattice Field Theory - LATTICE 2013\\
		July 29 - August 3, 2013\\
		Mainz, Germany}

\begin{document}

\section{Introduction}
\label{sect:intro}

The purpose of this study is to find a gauge system which may
cause the electroweak (EW) symmetry breaking.
Nowadays, we know that such gauge dynamics must give rise to large
anomalous dimension $(\sim 1)$ over wide range of energy scale and
also need the spontaneous breakdown of chiral symmetry at lower scale.

We focus here on a series of $\rm SU(2)_C$ gauge theories.
As for $\rm SU(2)_C$ gauge theories, the system with two adjoint Dirac fermions
has been studied extensively using numerical methods since the first work
\cite{Catterall_etal:2007}.
More recent studies of running coupling constant calculation
\cite{Hietanen_etal:2009,DeGrand_etal:2011}
and scaling of the spectrum \cite{DelDebbio_etal:2010} suggest that  this
system has an IR-fixed point.

We investigate $\rm SU(2)_C$ gauge theories with $N_f$ Dirac fermions
in the fundamental representation.
$\rm SU(2)_C$ gauge theory is one of the ${\rm Sp}(N)$ gauge theories so that its
fundamental representation is pseudoreal and that the plausible pattern of
chiral symmetry breaking is ${\rm SU}(2N_f) \rightarrow {\rm Sp}(2N_f)$.
Therefore the chiral dynamics of two-color QCD is different from that of
three-color QCD.
From the point of view of the application to the dynamical realization of EW
symmetry breaking, the effective Higgs sector of $\rm SU(2)_C$ gauge system
is quite different from that of $\rm SU(3)_C$ gauge system.
This fact also motivates us to perform lattice simulations to grasp its
properties of nonperturbative dynamics such as spectra of bound states.

Study of the system with fundamental fermions
intended to search the conformal window has also been done.
The perturbative approach \cite{Ritbergen_etal:1997} suggests
$6\le N_f^{\rm crtl}\le8$.
Reference~\cite{Iwasaki_etal:2004} investigated the phase structure of Wilson fermions
and indicated $N_f=3$ system is conformal in the IR limit.
Afterwards, running gauge coupling constant has been calculated
nonperturbatively for two-color QCD and then the conformal signal was indicated
for the system with $N_f=6$ \cite{DelDebbio_etal:2011}, $N_f=8$
\cite{Ohki_etal:2010}, and $N_f=10$ \cite{Karavirta_etal:2012}, while the QCD-like
signal was indicated for the system with $N_f = 4$ by Ref. \cite{Karavirta_etal:2012}.

We focus on the chiral properties of $\rm SU(2)_C$ gauge theory
with six Dirac fermions in the fundamental representation and show the result for
the quark mass dependence and the volume dependence of some quantities, such
as the mass of the lightest pseudoscalar meson, and decay constant.
Here we especially pay attention to the finite size effect on these quantities.
We carefully make sure which data do not suffer the finite size effect to extract
data which give the values in the infinite volume system and then compare these
results with the predictions from the theory with the breakdown of chiral symmetry
(Sec.~\ref{sect:mpi}, \ref{sect:cond}).
In some cases, we conversely use the finite size effect
that should reflect the long distance dynamics, and compare the obtained
data with the expectations in the system with the breakdown
of chiral symmetry (Sec.~\ref{sect:fpi}).

In this research, we employ the plaquette gauge action and the Wilson fermion
action on the lattices where spacial sizes are $16, 24$, and $32$.
In order to see the phase structure as a statistical system and fix a
parameter $\beta$, which is inversely proportional to the square of the bare coupling, we
investigate the hopping parameter dependence of the plaquette for several $\beta$.
We choose $\beta = 2.0$ where the bulk phase transition does not seem to
occur.

In the following sections, we give three signatures supporting that the six-flavor
theory is not the one with chiral symmetry breaking.
Here we report the essential points of them.
The full qualitative discussion is given in the full paper \cite{OurPaper:2013}.

\section{Pseudoscalar Spectrum}
\label{sect:mpi}
\begin{figure}[bh]
\vspace{-0.5mm}
\begin{tabular}{cc}
\begin{minipage}{0.47\hsize}
\begin{center}
\includegraphics[width=\hsize,clip]{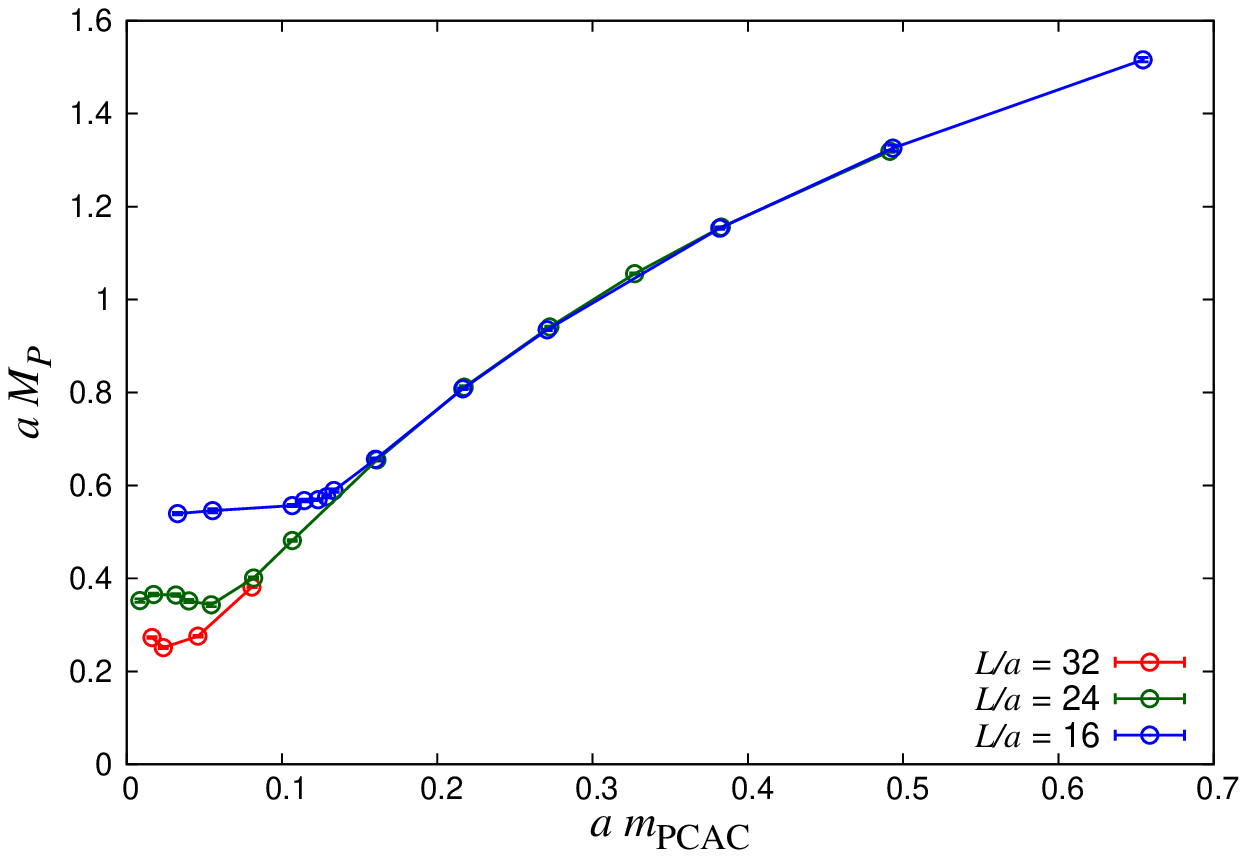}
\caption{
 The lightest pseudoscalar meson mass $aM_P$ versus $am_{\rm PCAC}$
in six-flavor theory.
}
\label{fig:mpi}
\end{center}
\end{minipage}
\quad
\begin{minipage}{0.47\hsize}
\begin{center}
\includegraphics[width=\hsize,clip]{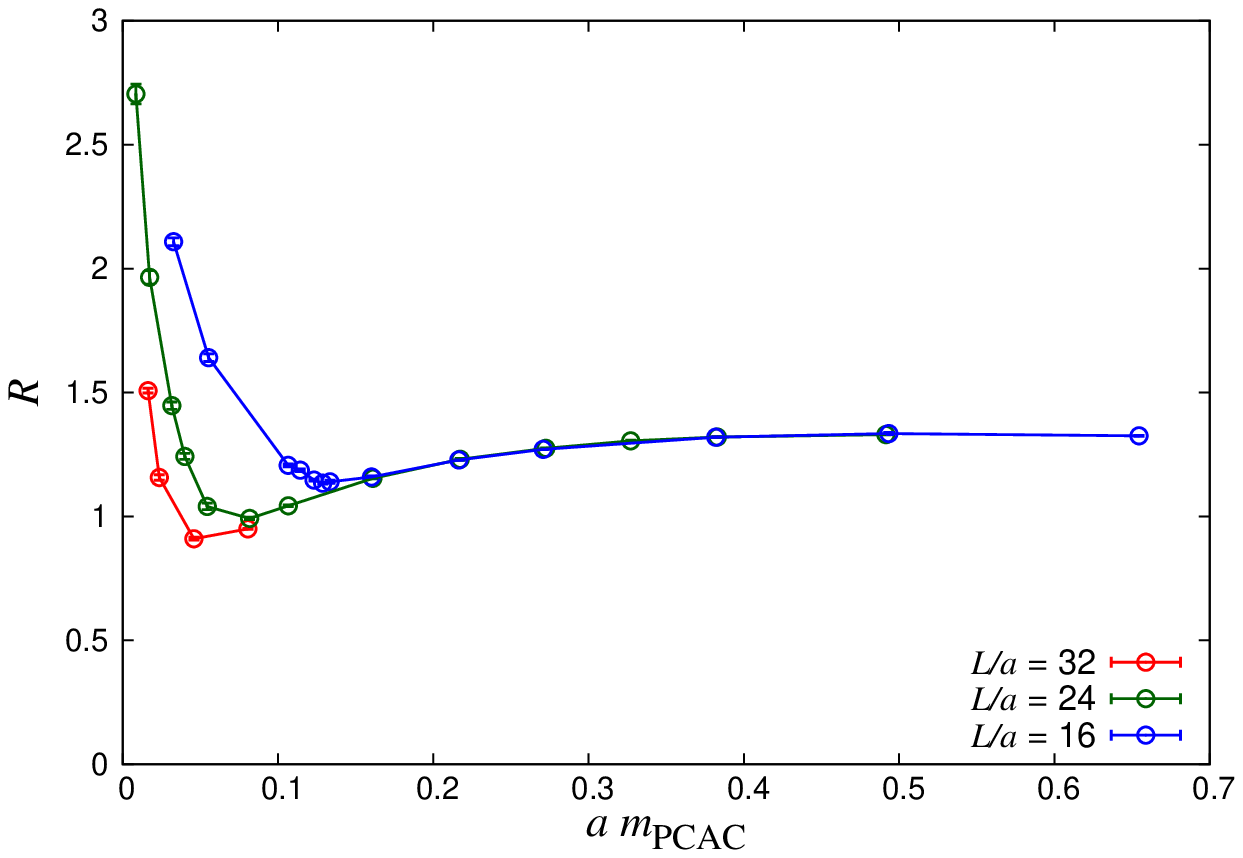}
\caption{
 The ratio $R=a^{1/2}M_P/{m_{\rm PCAC}}^{1/2}$ versus $am_{\rm PCAC}$
in six-flavor theory.
}
\label{fig:mpmq0.5}
\end{center}
\end{minipage}
\end{tabular}
\end{figure}

First, we discuss the mass of the lightest pseudoscalar channel: $M_P$.
Figure~\ref{fig:mpi} shows the quark mass dependence of $M_P$.
$M_P$ in $L/a=16$ and that in $L/a=24$ are consistent at
$am_{\rm PCAC}>0.2$.
For smaller $m_{\rm PCAC}$, on the other hand, $M_P$ seems to be bounded
from below and does not approach zero even in the chiral limit. The bound
on $M_P$ depends on the size of lattices in proportion to $\sim L^{-1}$,
and thus this seems to be caused by the finite size effect on $M_P$.
In order to judge whether the chiral symmetry breaking occurs in this system,
we analyze the quantity $R\equiv a^{1/2}M_P/{m_{\rm PCAC}}^{1/2}$
(Fig.~\ref{fig:mpmq0.5}).

Because of the finite size effect on $M_P$,
$R$ depends clearly on the system size for small quark masses
and diverges for $m_{\rm PCAC}\rightarrow0$.
In this observation, we neglect data with visible finite size effect. Then,
we notice the most important point from Fig.~\ref{fig:mpmq0.5};
%$M_P$ is not proportional to ${m_q}^{1/2}$
%for smaller quark mass region $am_{\rm PCAC} \lesssim 0.3$.
$M_P$ is proportional to ${m_{\rm PCAC}}^{1/2}$ for
$0.35<am_{\rm PCAC}<0.65$, but it changes with a different exponent,
$M_P$ is proportional to ${m_{\rm PCAC}}^\alpha$ with $\alpha > 0.5$,
for smaller quark mass region $am_{\rm PCAC} \lesssim 0.3$.
If the chiral symmetry breaking is realized in the system, $M_P$ should be
better approximated by ${m_q}^{1/2}$ for lighter quarks,
which is not acturally the case.
This is the first evidence supporting the
absence of chiral symmetry breaking in the six-flavor theory.

%%%%%%%%%%%%%%%%%%%%%%%%%%%%%%%%%%%%%
%%%%%%%%%%%%%%%%%%%%%%%%%%%%%%%%%%%%%
\vspace{-3mm}
%%%%%%%%%%%%%%%%%%%%%%%%%%%%%%%%%%%%%
\begin{figure}[bh]
\begin{tabular}{cc}
\begin{minipage}{0.47\hsize}
\begin{center}
\includegraphics[width=\hsize,clip]{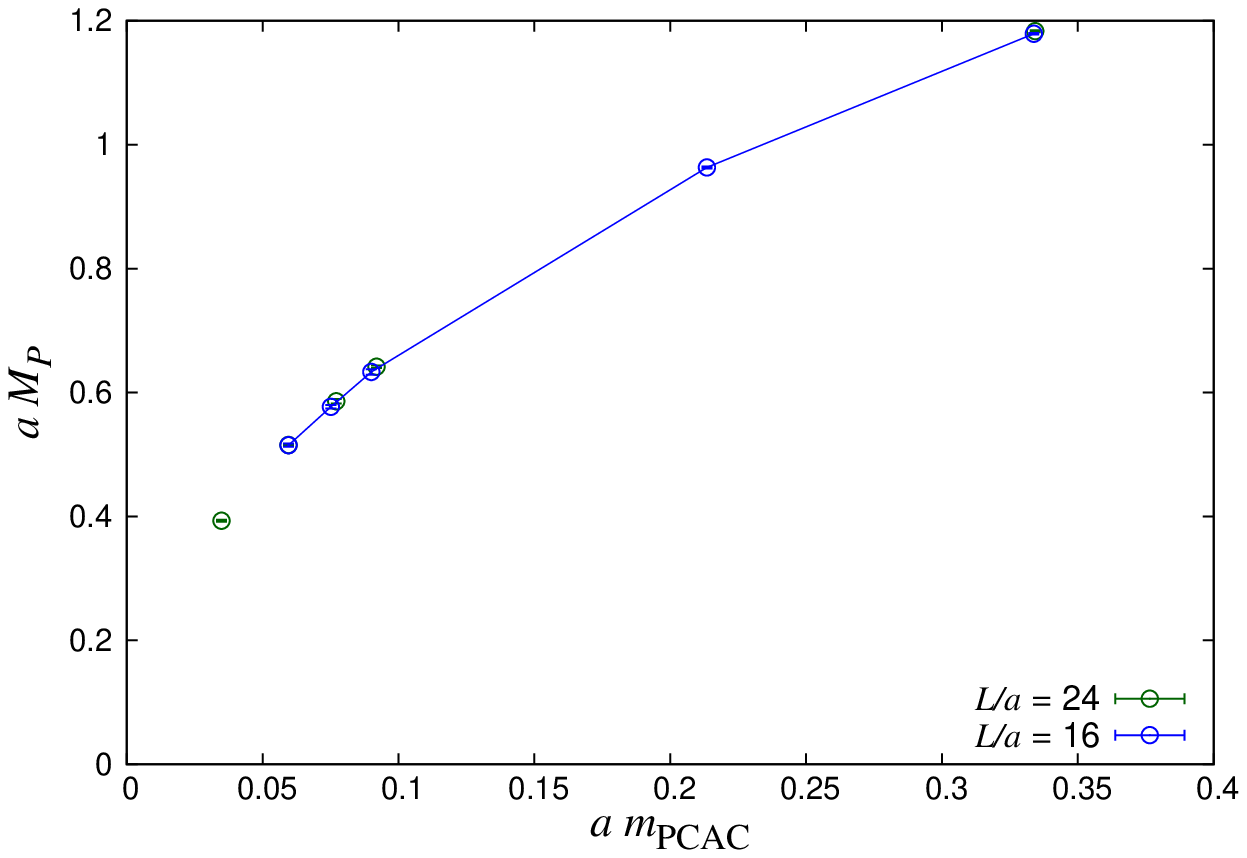}
\caption{
 The lightest pseudoscalar meson mass $aM_P$ versus $am_{\rm PCAC}$
in two-flavor theory.
}
\label{fig:mpi_Nf2}
\end{center}
\end{minipage}
\quad
\begin{minipage}{0.47\hsize}
\begin{center}
\includegraphics[width=\hsize,clip]{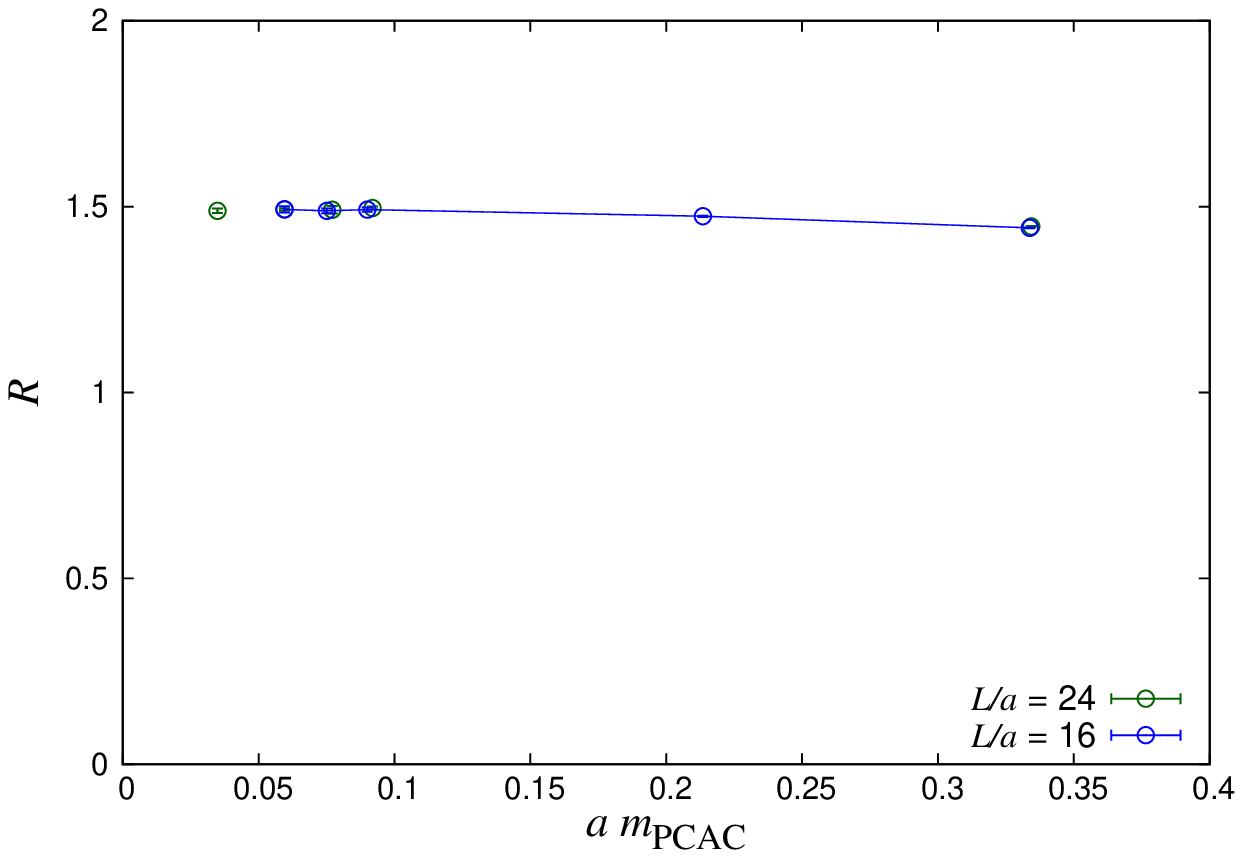}
\caption{
 The ratio $R=a^{1/2}M_P/{m_{\rm PCAC}}^{1/2}$ versus $am_{\rm PCAC}$
in two-flavor theory.
}
\label{fig:mpmq0.5_Nf2}
\end{center}
\end{minipage}
\end{tabular}
\end{figure}
%%%%%%%%%%%%%%%%%%%%%%%%%%%%%%%%%%%%%

In order to make this evidence more determinate, we compare this result with
the result in the two-flavor system.
In the simulation of the two-flavor theory, we employ lattices with sizes
$L/a=16,24$ and the same $\beta (=2.0)$ as in the six-flavor theory.
Figure~\ref{fig:mpi_Nf2} shows the quark mass dependence of $M_P$
in two-flavor theory, where no visible finite size effect on $M_P$ is seen.
Figure~\ref{fig:mpmq0.5_Nf2} shows the quark mass dependence of $R$,
where $M_P$ is better approximated by ${m_{\rm PCAC}}^{1/2}$
for smaller $m_{\rm PCAC}$. This behavior is compatible with the behavior
in the system with chiral symmetry breaking and able to be distinguished clearly
from the behavior in the case of $N_f=6$.

%      NEW SECTION
\section{Decay Constant}
\label{sect:fpi}
\begin{figure}[b]
\begin{tabular}{cc}
\begin{minipage}{0.47\hsize}
\begin{center}
\includegraphics[width=\hsize,clip]{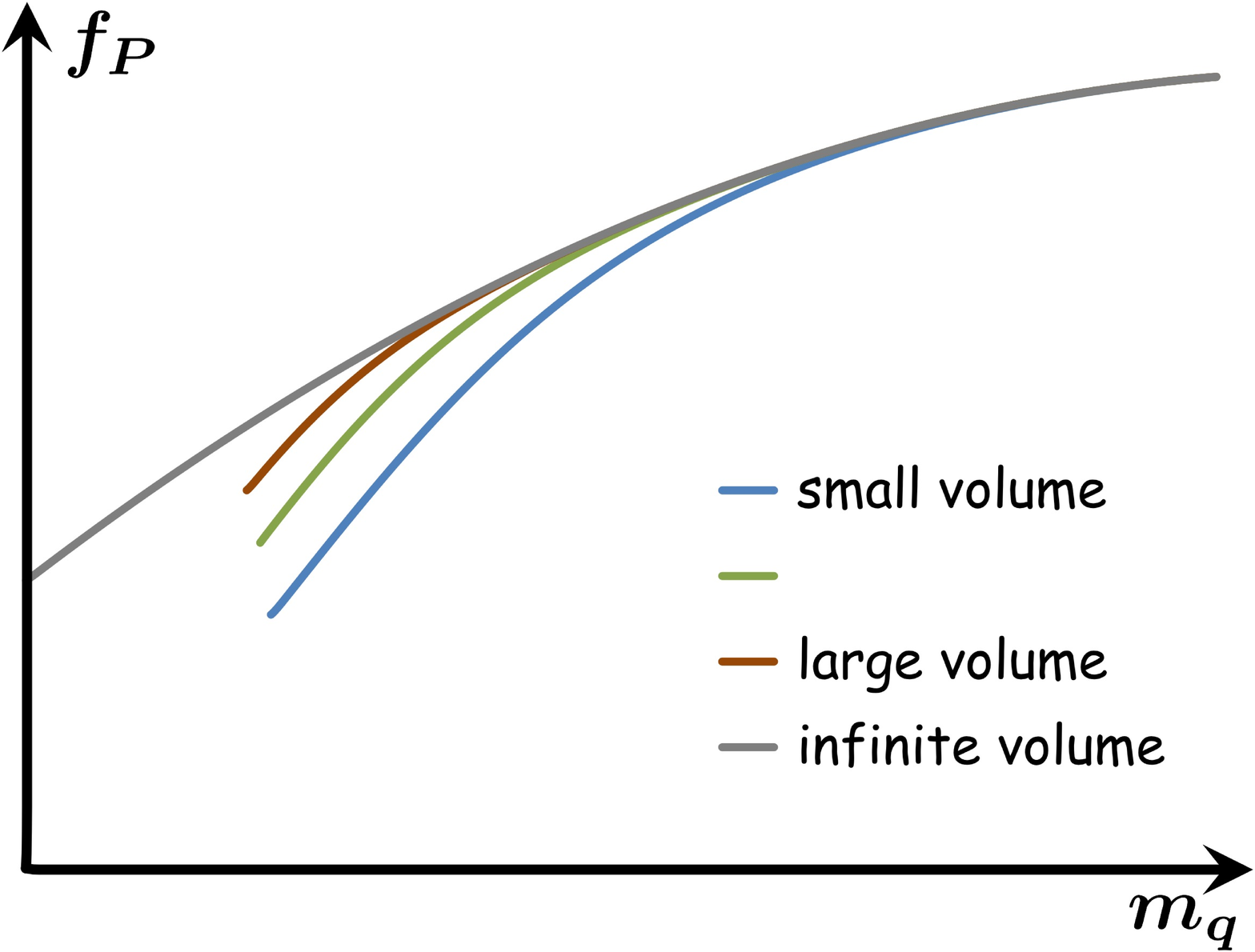}
\caption{
 A sketch of quark mass and volume dependence of $f_P$ in the system with
the breakdown of chiral symmetry. The finite size effect tends to decrease
$f_P$.
}
\label{fig:break_fpi}
\end{center}
\end{minipage}
\quad
\begin{minipage}{0.47\hsize}
\begin{center}
\vspace{2.5mm}
\includegraphics[width=\hsize,clip]{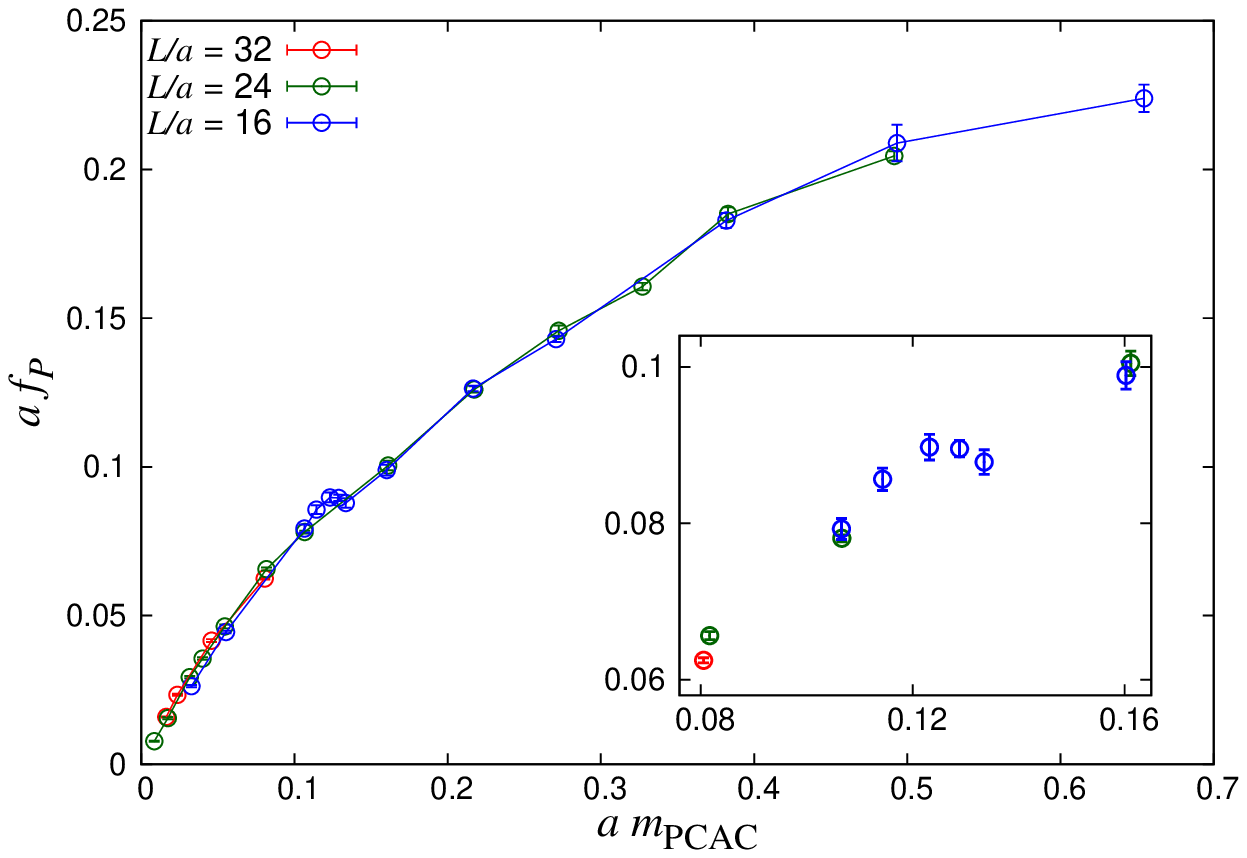}
\caption{
 The decay constant $af_P$ versus $am_{\rm PCAC}$ in six-flavor theory.
The finite size effect seems to increase $f_P$ around the region
$am_{\rm PCAC}=0.12\sim0.14$.
}
\label{fig:fpi}
\end{center}
\end{minipage}
\end{tabular}
\end{figure}

Next, we focus on the decay constant of the lightest pseudoscalar meson:
$f_P$. In the observation of this quantity, we especially pay attention to the
behavior of the finite size effect on $f_P$.
Figure~\ref{fig:break_fpi} shows a sketch of the quark mass and volume
dependence of $f_P$ which is theoretically expected for the system with the
breakdown of chiral symmetry.
The important point is that the finite size effect tends to decrease $f_P$.
This tendency is observed for both the $\epsilon$-regime
\cite{AokiFukaya2011} and p-regime \cite{Colangelo_etal2005}.
Therefore, if the finite size effect turns out to increase $f_P$,
it will give an evidence that
chiral symmetry breaking does not occur in the system.

Figure~\ref{fig:fpi} shows the quark mass dependence of $f_P$ defined by
the equation;
\begin{equation}
f_P
= 2\kappa\cdot 2m_{\rm PCAC} \sqrt{2A^{PP}\over M_P}
{1\over\sinh M_P},
\label{eq:fpi}
\end{equation}
which is derived from PCAC relation. In eq.~(\ref{eq:fpi}), $A^{PP}$ is
the amplitude of the PP correlator
\begin{equation}
\langle P(t)P(0)\rangle
\xrightarrow{\ a\ll t\ll T\ }
A^{PP}\big(e^{-M_Pt} + e^{-M_P (T-t)}\big).
\end{equation}

%%%%%%%%%%%%%%%%%%%%%%%%%%%%%%%%%%%%%
%%%%%%%%%%%%%%%%%%%%%%%%%%%%%%%%%%%%%

From Fig.~\ref{fig:fpi}, $f_P$ seems to vanish in the chiral limit.
When we pay attention to the definition (\ref{eq:fpi}), however, we notice that
this is not necessarily true because of the finite size effect on the pseudoscalar
meson mass $M_P$, which exists in the denominator of R.H.S. of eq. (\ref{eq:fpi})
while $m_{\rm PCAC}$ exists in the numerator.
Since $M_P$ at a finite volume system is bounded from below at small quark
mass region and cannot vanish in the chiral limit, $f_P$ defined as (\ref{eq:fpi})
at a finite volume system vanishes even if chiral symmetry breaking
occurs in this system.
Therefore, the result in the mass region where the finite size effect on $M_P$
dominates should be neglected.

Then, we notice that, for the lattice size $L/a=16$, there is the quark mass region
$am_{\rm PCAC}=0.12\sim0.14$ where the finite size effect seems to increase
$f_P$.
The behavior of this finite size effect is opposite to that in the system with
chiral symmetry breaking.
This is the second signature supporting that the chiral symmetry breaking
does not occur in the six-flavor theory.

\section{Chiral Condensate}
\label{sect:cond}
Lastly, we discuss the chiral condensate $\langle\overline\psi\psi\rangle$,
which is the order parameter of chiral symmetry breaking.
Since we use the Wilson fermion, however, $\langle\overline\psi\psi\rangle$ is
dominated by the UV divergence $\sim O(a^{-3})$, which does not vanish in
the chiral limit.
Instead of $\langle\overline\psi\psi\rangle$, we focus on the {\it subtracted}
chiral condensate $\langle\overline\psi\psi\rangle_{\rm subt}$, which is free
from such a cubic divergence.
The subtracted chiral condensate in the Wilson fermion is calculated by using
the Ward-Takahashi identity with respect to the axial-vector current
\cite{Bochicchio_etal1985};
\begin{equation}
\langle\overline\psi\psi\rangle_{\rm subt}
= 2m_{\rm PCAC}\cdot(2\kappa)^2\sum_n
\langle P(n)P(0)\rangle.
\end{equation}

\begin{figure}[htp]
\begin{tabular}{cc}
\begin{minipage}{0.47\hsize}
\begin{center}
\vspace{2.2mm}
\includegraphics[width=\hsize,clip]{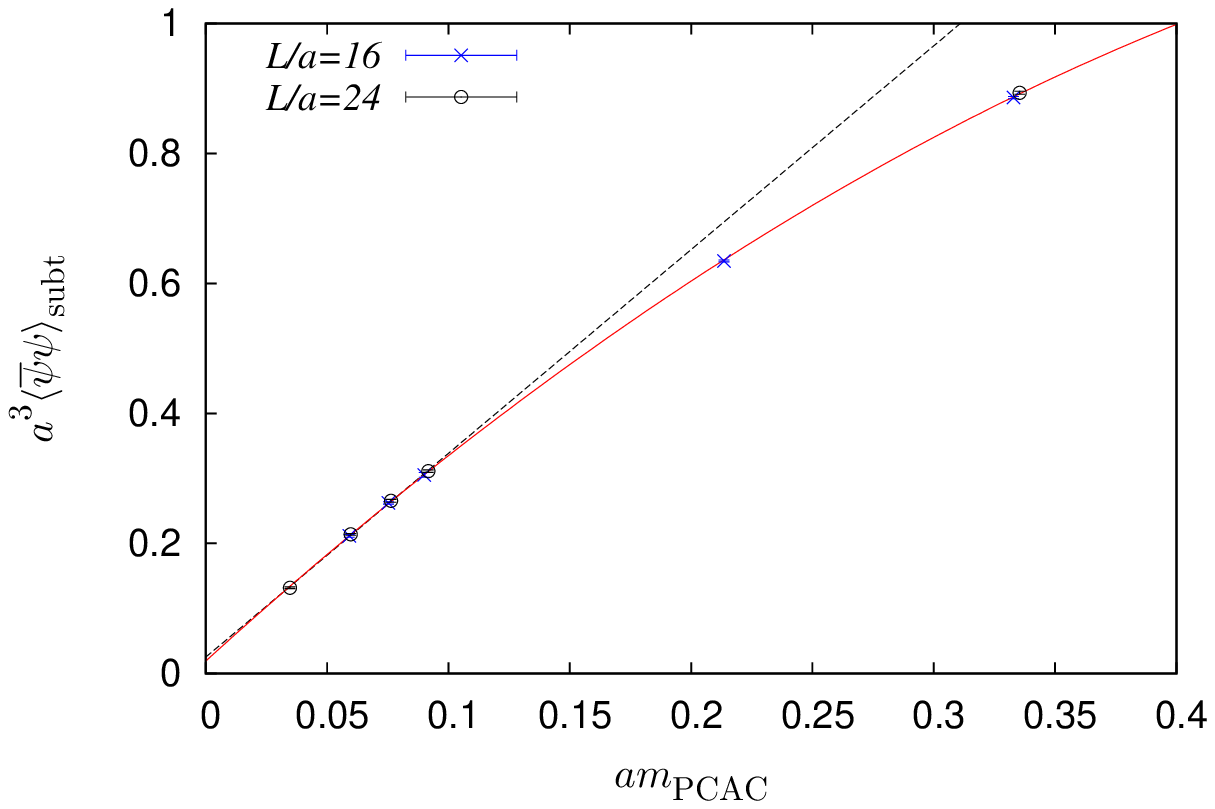}
\caption{
 The subtracted chiral condensate
$\langle\overline\psi\psi\rangle_{\rm subt}$ versus $am_{\rm PCAC}$
in two-flavor theory with linear(dotted line) and quadratic(red line) fit lines.
}
\label{fig:cond_Nf2}
\end{center}
\end{minipage}
\quad
\begin{minipage}{0.47\hsize}
\begin{center}
\includegraphics[width=\hsize,clip]{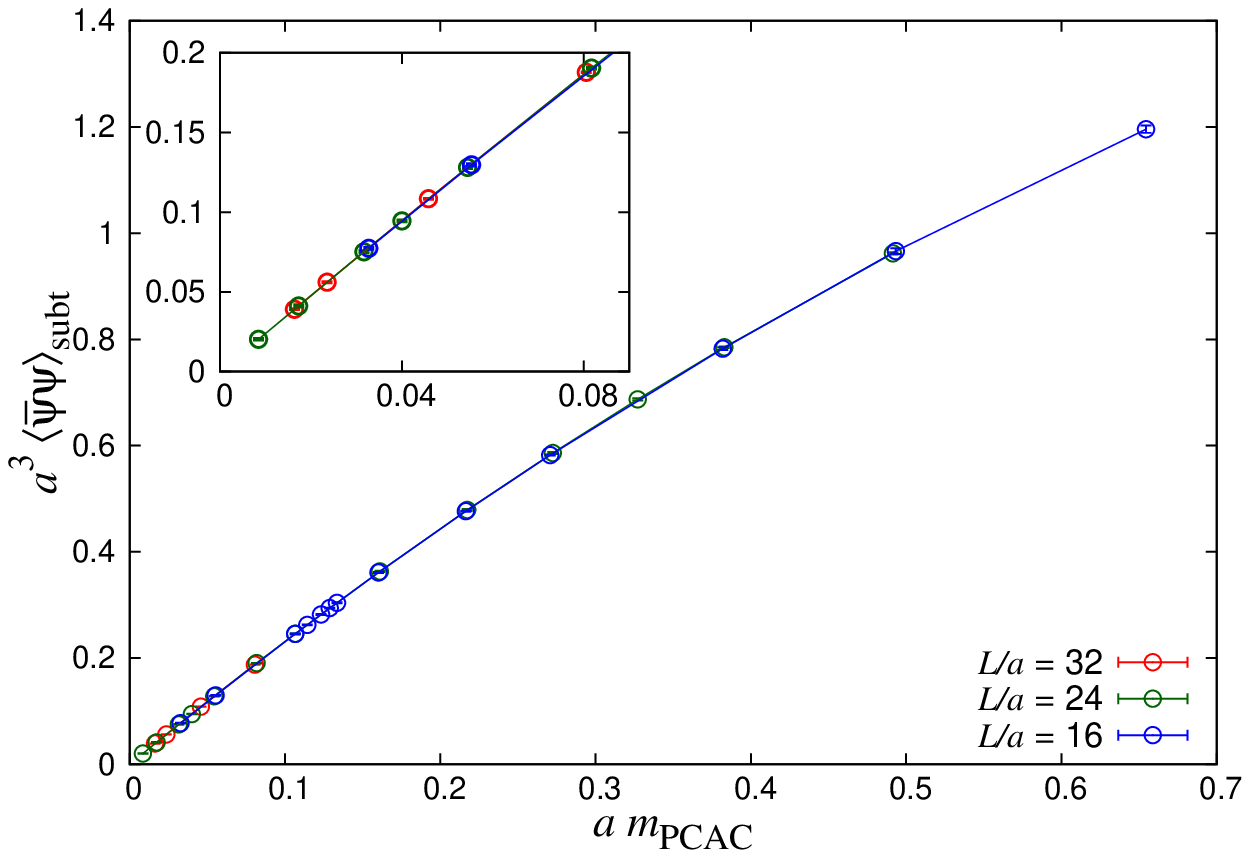}
\caption{
 The subtracted chiral condensate
$\langle\overline\psi\psi\rangle_{\rm subt}$ versus $am_{\rm PCAC}$
in six-flavor theory.
}
\label{fig:cond}
\end{center}
\end{minipage}
\end{tabular}
\end{figure}

Before discussing the result in the six-flavor theory, we will discuss the utility of the
subtracted chiral condensate by analyzing the result in the two-flavor theory.
Figure~\ref{fig:cond_Nf2} shows the quark mass dependence of
$\langle\overline\psi\psi\rangle_{\rm subt}$ in the two-flavor theory with two
fit lines.
Since there is no visible volume dependence on 
$\langle\overline\psi\psi\rangle_{\rm subt}$, we assume these data do not
suffer the finite size effect and results of these fit give the value in the infinite
volume system.
The dotted line is drown by fitting the linear function
\begin{equation}
f_2(x=am_{\rm PCAC}) = a_0 + a_1x
\label{eq:linear_func}
\end{equation}
to the data of $am_{\rm PCAC}<0.1$,
while the red line is drown also by fitting the quadratic function
\begin{equation}
f_3(x=am_{\rm PCAC}) = b_0 + b_1x + b_2x^2
\label{eq:quadratic_func}
\end{equation}
to all data. Then, we find
\begin{equation}
\lim_{m_{\rm PCAC}\rightarrow0}
a^3 \langle\overline\psi\psi\rangle_{\rm subt}\big|_{N_f=2}
= \Bigg\{
\begin{array}{ll}
0.0230(22) & {\rm linear\ fit}
\\[1.8mm]
0.0185(17) & {\rm quadratic\ fit}
\end{array}
.
\end{equation}
Both of these values indicate that the chiral limit of
$\langle\overline\psi\psi\rangle_{\rm subt}$ is nonzero and much smaller
than $a^{-3}$ so we assume the chirally extrapolated value of
$\langle\overline\psi\psi\rangle_{\rm subt}$ is free from the UV divergence.
Therefore, we calculate $\langle\overline\psi\psi\rangle_{\rm subt}$ also
in the six-flavor theory.

%%%%%%%%%%%%%%%%%%%%%%%%%%%%%%%%%%%%%
%%%%%%%%%%%%%%%%%%%%%%%%%%%%%%%%%%%%%

Figure~\ref{fig:cond} shows the quark mass dependence of
$\langle\overline\psi\psi\rangle_{\rm subt}$ in the six-flavor theory.
We perform the chiral extrapolation also for this result to judge
whether the extrapolated value is consistent with zero within available precision,
i.e. whether the result supports two signatures gained from the results of $M_P$
and $f_P$, which indicate that chiral symmetry breaking does not occur in this
system.
There is no visible volume dependence on calculated values of
$\langle\overline\psi\psi\rangle_{\rm subt}$.
Therefore we assume these data suffer few finite size effect so
that we can use the value at the lightest quark mass as one of target data
to be fit.
We introduce the result of the quadratic fit (\ref{eq:quadratic_func}) using
following two data sets;
\\[2mm]\hspace{7mm}
{\bf (S1):} data with $am_{\rm PCAC}\le0.2$,
\\[2mm]\hspace{7mm}
{\bf (S2):} data with $am_{\rm PCAC}\le0.1$.
\\[2mm]
The fit gives
\begin{equation}
\lim_{m_{\rm PCAC}\rightarrow 0}
a^3\langle\overline\psi\psi\rangle_{\rm subt}\big|_{N_f=6}
= \Bigg\{
\begin{array}{rll}
-0.00001(27) &\ & {\rm data\ set:\ \bf S1}
\\[1.8mm]
0.00012(41) &\ & {\rm data\ set:\ \bf S2}
\end{array}
,
\end{equation}
which are vanishing extrapolated values within available precision.
This is the third signature supporting the absence of the breakdown of
chiral symmetry.

Meanwhile, we also perform the power type fit;
\begin{equation}
f_{\rm pow}(x=am_{\rm PCAC}) = c_1 x + c_2 x^\alpha,
\end{equation}
which is expected for small quark masses in the system with IR-fixed point.
If the system is conformal in the infrared and the hyper-scaling hypothesis
is valid, $\alpha$ is equal to $(3-\gamma_\star) / (1 + \gamma_\star)$,
where $\gamma_\star$ denotes the mass anomalous dimension at the IR-fixed
point
\cite{DelDebbio_etal:2010hyp}.
As a result, we obtain $\alpha = 1.76(66)$ for the data set {\bf S2}.
This indicates
\begin{equation}
0.17\le \gamma_\star \le 0.90,
\end{equation}
which is not inconsistent with that found in the
study of the running coupling constant defined in the Schr\"{o}dinger functional
scheme \cite{Hayakawa_etal:2013};
\begin{equation}
0.26 \le \gamma_{\star,\rm SF} \le 0.74.
\end{equation}

\section{Summary}

We investigated the chiral behavior of the two-color QCD with six fundamental
Dirac fermions by analyzing the quark mass dependence and the volume
dependence of some quantities with very careful attention to the finite size effects
on these quantities.
Since we observe that the presented three quantities behave oppositely to the
prediction from the theory with chiral symmetry breaking,
we conclude the breakdown of chiral symmetry does not occur in this
gauge system.

More detailed and more quantitative discussions are summarized in the full
paper \cite{OurPaper:2013}.
The same conclusion is obtained in the calculation of the SF running coupling
constant reported in the companion paper \cite{Hayakawa_etal:2013}.

\begin{acknowledgments}
 The numerical simulations were 
carried out on the computer system $\varphi$ at Nagoya University,
and the servers equipped with GPU cards
at High Energy Accelerator Research Organization (KEK).
 This work is supported partly by JSPS Grands-in-Aid
for Scientific Research 20540261, 22224003, 22740183, 23740177.
\end{acknowledgments}

\end{document}